\documentclass{article}
\usepackage{spconf,amsmath,graphicx}
\usepackage{booktabs}
\usepackage{multirow}
\usepackage{subcaption}
\usepackage[colorlinks]{hyperref}
\usepackage{amssymb}
\usepackage{algorithm}
\usepackage{algpseudocode}
\hypersetup{
    breaklinks=true,   
    colorlinks=true,   
    pdfusetitle=true,  
}

\usepackage{enumitem}
\setlist{nosep, leftmargin=14pt}

\usepackage{mwe} 

\title{Learning Cortical Anomaly through Masked Encoding for Unsupervised Heterogeneity Mapping}
\name{%
\begin{tabular}{@{}c@{}}
Hao-Chun Yang$^{13}$\qquad
Ole Andreassen$^{2}$\qquad
Lars Tjelta Westlye$^{2}$\\
Andre F. Marquand$^{3}$\qquad
Christian F. Beckmann$^{3}$\qquad
Thomas Wolfers$^{123}$
\end{tabular}}

\address{
    $^{1}$ \normalsize{Department of Psychiatry and Psychotherapy, Tübingen Center for Mental Health (TÜCMH), Germany}\\
    $^{2}$ \normalsize{Norwegian Centre for Mental Disorders Research, Division of Mental Health and Addiction, Norway}\\
    $^{3}$ \normalsize{Donders Institute for Brain, Cognition, and Behaviour, Radboud University, Netherlands}
}
\begin{document}
\maketitle

\begin{abstract}
    The detection of heterogeneous mental disorders based on brain readouts remains challenging due to the complexity of symptoms and the absence of reliable biomarkers. This paper introduces CAM (\underline{C}ortical \underline{A}nomaly Detection through \underline{M}asked Image Modeling), a novel self-supervised framework designed for the unsupervised detection of complex brain disorders using cortical surface features. We employ this framework for the detection of individuals on the psychotic spectrum and demonstrate its capabilities compared to state-of-the-art methods, achieving an AUC of $0.696$ for Schizoaffective and $0.769$ for Schizophreniform, without the need for any labels. Furthermore, the analysis of atypical cortical regions, including \textit{Pars Triangularis} and several frontal areas often implicated in schizophrenia, provides further confidence in our approach. Altogether, we demonstrate a scalable approach for anomaly detection of complex brain disorders based on cortical abnormalities. The code will be made available at \url{https://github.com/chadHGY/CAM}.
\end{abstract}

\section{Introduction}
\label{sec:intro}
The detection of mental disorders through brain imaging has garnered significant attention in recent years. Current diagnostic methods heavily rely on self-reported experiences, clinical observations, and the exclusion of alternative explanations for symptom emergence. However, the dynamic nature of these illnesses, coupled with the lack of objective diagnostic tests, often leads to misdiagnosis \cite{merten2017overdiagnosis, gara2019naturalistic}. Therefore, the development of reliable quantitative biomarkers is crucial to facilitate early detection of emerging illnesses.

Most existing methods for detecting mental disorders rely on supervised learning approaches applied to neuroimaging data, such as 3D brain MR volumes or functional MRI \cite{oh2020identifying, zhang2023detecting, bi2023multivit}. Nevertheless, these methods have limitations. Firstly, supervised learning requires large matched patient and control datasets, which can be challenging and costly to obtain, and does not fully account for the heterogeneous manifestations and overlapping symptoms across different mental disorders \cite{bzdok2018machine, wolfers2021replicating}. Secondly, most machine learning and deep learning methods primarily focus on volumetric or functional modalities, with limited exploration of cortical surface features. The cortical surface exhibits intricate morphological patterns that could provide insights into brain structure and function \cite{song2022linking, dahan2022surface}. Therefore, leveraging an unsupervised approach to model high-dimensional cortical surface data could facilitate the detection of brain anomalies, aiding in the identification of novel illness subtypes and enhancing generalizability to new data and populations without relying on labeled supervision.

In this study, we propose an analytical framework, CAM (Cortical Anomaly Detection through Masked Image Modeling), for unsupervised detection of complex brain disorders using cortical surface features. We utilize a pretext task of masked image modeling to learn representations of cortical surface features in a self-supervised manner and employ a novel iterative masked anomaly detection algorithm to discover deviations from those learned representations. We validate our approach in disorders on the psychotic spectrum \cite{skaatun2016global}, including Schizophrenia (SZ), Bipolar Disorder (BD), Schizoaffective (SA), and Schizophreniform (SZF), which have been linked to cortical alterations previously \cite{wolfers2021replicating}. Our experiments reveal that CAM can be used to detect cortical anomalies in individuals on the psychotic spectrum. The primary contributions of this work can be summarized as follows:

\begin{figure*}[ht!]
    \centering
    \includegraphics[width=\textwidth]{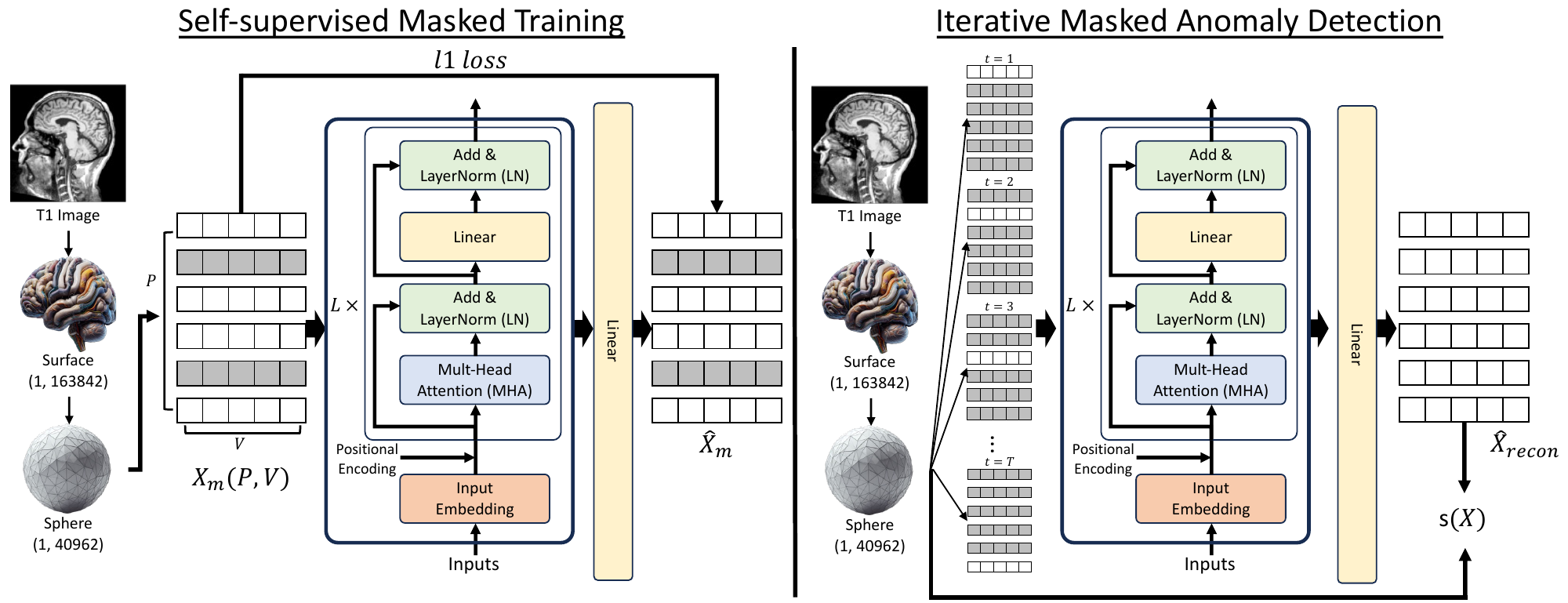}
    \caption{Our proposed \underline{C}ortical \underline{A}nomaly detection through \underline{M}asked image modeling (CAM) framework.}
    \label{fig:flowchart}
\end{figure*}
\begin{itemize}
    \item We introduced a novel method that learns high-dimensional abstractions from cortical features for unsupervised brain anomaly detection.
    \item We demonstrated the effectiveness of CAM in distinguishing psychotic disorders from healthy controls.
    \item We highlighted CAM's capabilities in identifying key cortical regions involved in psychotic disorders, providing additional validation and confidence in our approach.
\end{itemize}

\section{Methods}
\label{sec:methods}
Inspired by the method proposed by Dahan et al. \cite{dahan2022surface}, we extracted surface features from raw T1 images and resampled them into a sequence of patches. These patches were then input into our CAM framework, which comprises a self-supervised vision transformer encoder for masked patch prediction, followed by an iterative masked prediction approach where the trained encoder is used for unsupervised anomaly detection. Further details are elaborated in the subsequent sections.

\subsection{Data Preprocessing}
\label{sec:preprocessing}
We extracted surface features from T1 MRI images using the Freesurfer \textit{recon-all} pipeline (version 6.0) with default parameters. This process generated six mesh features: \textit{Area}, \textit{Curvature}, \textit{Inflated surface}, \textit{Sulc}, \textit{Thickness}, and \textit{Volume}, each corresponding to the Desikan-Killiany atlas \cite{fischl2012freesurfer}. Subsequently, we re-tessellated all individual meshes using barycentric interpolation, from their template resolution (163,842 vertices) to a sixth-order icosphere (40,962 equally spaced vertices). Finally, the icosphere was divided into triangular patches of equal vertex count ($P$, $V$) = (320, 153) covering the entire spherical space, where $P$ represented the number of patches, and $V$ represented the number of vertices per patch for the subsequent Transformer encoder. It's noteworthy that our method did not require further inter-subject registration nor any atlas-based division.

\begin{algorithm}
      \caption{Iterative Masked Anomaly Detection}
      \label{algo:masked_anomaly_detection}
      \begin{algorithmic}[1]
            \Require Data samples $\mathbf{X}$, encoder $f_\theta$, iteration count $T$=10
            \Ensure Anomaly score $s(\mathbf{X})$.
            \State $\text{patch\_size} \gets \text{total number of patches in } \mathbf{X}$
            \State $\text{step} \gets \text{int}(\text{patch\_size} / T)$
            \State $\text{start} \gets 0$
            \State $\text{end} \gets \text{step}$
            \For{$t=1$ to $T$}
            \State Mask patches from $\text{start}$ to $\text{end}$ in $\mathbf{X}$ to obtain $\mathbf{X}^t_m$
            \State Obtain reconstructions: $\hat{\mathbf{x}}^t_m \gets g\phi(f_\theta(\mathbf{X}^t_m))$
            \State $\text{start} \gets \text{end}$
            \State $\text{end} \gets \text{start} + \text{step}$
            \EndFor
            \State $\hat{\mathbf{X}}_{\text{recon}} \gets [\hat{\mathbf{x}}_m^1, \hat{\mathbf{x}}_m^2, ..., \hat{\mathbf{x}}_m^{T}]$
            \State Compute the anomaly score: $s(\mathbf{X}) = ||\mathbf{X} - \hat{\mathbf{X}}_{\text{recon}}||_1$.
      \end{algorithmic}
\end{algorithm}
\subsection{Self-supervised Masked Training}
\label{sec:masked_training}
Given an input sequence $\mathbf{X} \in \mathbb{R}^{P \times V}$, where $P$ is the number of patches, and $V$ is the dimension of each patch, we perform self-supervised masked training as follows:
We first randomly mask out $M$ patches and replace them with learned mask tokens to obtain the masked input sequence $\mathbf{X}_m$. We then use a vision transformer encoder $f_\theta$, comprising $L$ self-attention layers, to extract features:
\begin{equation}
    \mathbf{Z}^{l+1} = f_\theta^l(\mathbf{Z}^l), \quad \mathbf{Z}^0 = \mathbf{X}_m
\end{equation}

\begin{equation}
    f_\theta^l(\mathbf{Z}^l) = \mathrm{LN}(\mathrm{MHA}(\mathrm{LN}(\mathbf{Z}^l)) + \mathbf{Z}^l)
\end{equation}
Here, $\mathrm{LN}$ represents layer normalization, and $\mathrm{MHA}$ denotes multi-headed self-attention \cite{xie2022simmim}. The output features $\mathbf{Z} = \mathbf{Z}^L \in \mathbb{R}^{P \times D}$, where $D$ is the hidden feature dimension, are then used to predict the masked patches through a linear layer $g_{\phi}$:
\begin{equation}
    \hat{\mathbf{X}}_m = g_{\phi}(\mathbf{Z}_m), \forall m \in M
\end{equation}
where $\mathbf{Z}_m$ denotes the feature vector corresponding to the $m$-th masked patch. Finally, the parameters $\theta$ and $\phi$ are trained by minimizing the $\ell_1$ loss between the predicted and ground-truth masked patches:
\begin{equation}
    \mathcal{L}(\theta, \phi) = \frac{1}{|M|} \sum_{m \in M} | \hat{\mathbf{X}}_m - \mathbf{X}_m |_1
\end{equation}
This framework encourages the model to make predictions based on spherical-spatial contextual information, facilitating the learning of complex natural patterns among brain cortical regions.

\subsection{Iterative Masked Anomaly Detection}
After pre-training the encoder $f_\theta$, we applied it for unsupervised anomaly detection on new data samples $\mathbf{X}$. We proposed an iterative masked prediction approach, as outlined in Algorithm \ref{algo:masked_anomaly_detection}. This algorithm iteratively masks a certain percentage of patches in the input data $\mathbf{X}$ and reconstructs the masked patches based on the context of other visible patches. The final anomaly score $s(\mathbf{X})$ is computed as the $\ell_1$ distance between the original data and the reconstructed data. This score is then averaged across the atlas's regions of interest (ROIs). A higher score indicates a higher likelihood of the sample being anomalous compared to the training (healthy) subjects.
\section{Experiments}\label{sec:experiment}

\subsection{Experimental Setup}\label{ssec:setup}
\begin{table}[h]
    \centering
    \caption{Demographics of the experimental data. OPN is used for training and validation, while TOP is used for testing unsupervised anomaly detection.}
    \resizebox{0.95\columnwidth}{!}{%
        \begin{tabular}{@{}cccc@{}}
            \toprule
            \multicolumn{4}{c}{\textbf{\normalsize{OPN}} \cite{OpenNeuro2021}}     \\ \midrule\midrule
            Group                  & No.Subjects & Age Range & Sex (M/F)           \\
            Healthy Control (HC)   & 1135        & 5-73      & 44\%/56\%           \\ \toprule
            \multicolumn{4}{c}{\textbf{\normalsize{TOP}} \cite{skaatun2016global}} \\ \midrule\midrule
            Group                  & No.Subjects & Age Range & Sex (M/F)           \\
            Healthy Control (HC)   & 290         & 18-59     & 54\%/46\%           \\
            Schizophrenia (SZ)     & 165         & 19-60     & 65\%/35\%           \\
            Bipolar Disorder (BD)  & 189         & 17-65     & 42\%/58\%           \\
            Schizoaffective (SA)   & 33          & 20-62     & 30\%/70\%           \\
            Schizophreniform (SZF) & 22          & 19-45     & 50\%/50\%           \\ \bottomrule
        \end{tabular}
    }
    \label{table:demographics}
\end{table}

\subsubsection{Datasets}\label{sssec:datasets}
Table \ref{table:demographics} summarizes the demographics of the dataset. Subjects with a median-centered absolute Euler number greater than 25 were excluded, as these were found to be of poor quality \cite{kia2022closing}. The OPN dataset was divided in a 60/40 ratio with balanced age/gender for training and validation, while the TOP dataset was reserved for testing. Considering the geometric symmetry of the brain surface, our initial exploration focused on left-hemisphere features throughout the experiments.

\subsubsection{Comparison Methods}\label{sssec:comparison_methods}
To demonstrate the effectiveness of surface features for unsupervised schizophrenia-spectrum detection, we compared our method with standard statistical features, namely \textit{aparc.a2009s.stats} \cite{fischl2012freesurfer}, using a series of models (ABOD \cite{abod2008}, IForest \cite{iforest2008}, GMM \cite{gmm2015}, ECOD \cite{ecod2022}) as baselines. Additionally, we compared our method with state-of-the-art deep learning methods VAE \cite{vae2020} and DAE \cite{dae2022}, which have demonstrated superior performance in unsupervised anomaly detection tasks. Both VAE and DAE were adapted to accept statistical feature input by replacing the convolutional layers with linear layers. Anomaly scores were then averaged across ROIs for a fair comparison.

\subsubsection{Training and Evaluation Details}\label{sssec:training_and_evaluation}
All methods, including baseline models and our CAM framework, were trained on the OPN training set and underwent hyperparameter searches using the OPN validation set. The best model was then evaluated on the hold-out TOP testing set to prevent information leakage. For non-deep-learning models, hyperparameters were searched according to the original papers: ABOD (1-25 neighbors), IForest (50-200 estimators), and GMM (1-25 components). VAE and DAE had encoder architectures chosen from [[128, 64, 32], [256, 128, 64], [512, 256, 128]] with reversed decoder layers, and the latent dimension from [16, 32, 64] with LeakyReLU activation. For CAM, the specific hyperparameters used were: $L$ (number of transformer layers) = 6, $H$ (number of attention heads) = 12, $D$ (latent dimension) = 64, and $M$ (number of masked patches in training) = 50\% of the total patches, as recommended by \cite{xie2022simmim}. All neural networks (VAE, DAE, CAM) were trained using established techniques, including: AdamW optimization with a learning rate selected from [1$e$-3, 1$e$-5], a cosine annealing scheduler, and early stopping with a maximum of 200 epochs. The $\ell_1$ loss function was utilized, and the performance was assessed using the area under the receiver operating characteristic curve (AUC) for the best ROIs.

\begin{table*}[th]
    \centering
    \Large
    \caption{Unsupervised anomaly detection results. AUC is the evaluation metric, where chance=0.5. Bold indicates the best results for each disease group. Features: A (Area), C (Curvature), I (Inflated surface), S (Sulc), T (Thickness), V (Volume). A permutation test with 10,000 permutations was conducted to assess the likelihood of the measured AUC being due to chance. *p-value $<$ 0.05, **p-value $<$ 0.01}
    \resizebox{\textwidth}{!}{%
        \begin{tabular}{@{}ccccc|cc|cccccc@{}}
            \toprule
                      & ABOD\cite{abod2008} & IForest\cite{iforest2008} & GMM\cite{gmm2015} & ECOD\cite{ecod2022} & VAE\cite{vae2020} & DAE\cite{dae2022} & CAM(A) & CAM(C) & CAM(I) & CAM(S) & CAM(T)           & CAM(V)  \\ \midrule\midrule
            HC vs SZ  & 0.529               & 0.545                     & 0.611*            & 0.563               & 0.597*            & 0.593             & 0.573  & 0.593  & 0.601* & 0.625* & \textbf{0.666*}  & 0.577   \\
            HC vs BD  & 0.535               & 0.497                     & 0.544             & 0.505               & 0.578             & 0.584             & 0.557  & 0.569  & 0.586  & 0.565  & \textbf{0.627}*  & 0.568   \\
            HC vs SA  & 0.523               & 0.607*                    & 0.580             & 0.621*              & 0.660*            & 0.662*            & 0.668* & 0.654* & 0.629* & 0.642* & \textbf{0.696}** & 0.668*  \\
            HC vs SZF & 0.580               & 0.472                     & 0.654*            & 0.654*              & 0.709**           & 0.697**           & 0.666* & 0.621* & 0.640* & 0.650* & \textbf{0.769}** & 0.698** \\ \bottomrule
        \end{tabular}
    }
    \label{table:anomaly_detection_results}
\end{table*}

\subsection{Unsupervised Anomaly Detection Results}\label{ssec:anomaly}
Table \ref{table:anomaly_detection_results} summarizes the unsupervised anomaly detection results. The deep learning models (VAE, DAE) consistently outperformed chance (AUC=0.5) compared to the baseline machine learning models (ABOD, IForest, GMM, ECOD), except for HC vs SZ. This was expected, as deep learning models can learn complex patterns among brain cortical regions. We then found that Surface features generally improved disease discrimination, especially cortical \textit{Thickness} (CAM(T)). In particular, CAM(T) achieved the best performance in distinguishing healthy controls from individuals with Schizophrenia (HC vs SZ, AUC=0.666) Schizoaffective (HC vs SA, AUC=0.696) and Schizophreniform (HC vs SZF, AUC=0.769) disorders. This suggests surface features contain inherently discriminative information for psychotic spectrum disorders, which our CAM framework effectively learned.

\vspace{-2mm}
\subsection{Key ROIs Analysis}\label{ssec:key_rois}
\vspace{-4mm}
\begin{table}[ht]
    \centering
    \caption{ROIs identified by two-tailed Student’s t-test (p-value $<0.01$) utilizing CAM(T)'s anomaly scores. We also report the measured AUC of each region.}
    \resizebox{\columnwidth}{!}{%
        \begin{tabular}{@{}cc@{}}
            \toprule
                          & ROIs                      \\ \midrule
            ~~~~HC vs SA  & Pars Triangularis (0.696) \\
            \addlinespace 
            ~~~~HC vs SZF & \begin{tabular}[c]{@{}c@{}}Superior Frontal (0.769),\\ Rostral Middle Frontal (0.750),\\ ~~~~~~~~~Frontal Pole (0.693), Parsorbitalis (0.660)~~~~~~~~~\end{tabular} \\
            \bottomrule
        \end{tabular}
    }
    \label{table:key_rois}
    \vspace{-2mm}
\end{table}
To identify key ROIs contributing to disease discrimination, we conducted a two-tailed Student's t-test on the CAM(T) anomaly scores between the healthy control (HC) group and each disease group. As summarized in Table \ref{table:key_rois}, the identified significant ROIs were consistent with previous findings. For example, the involvement of \textit{Pars Triangularis} in Schizoaffective (HC vs SA) \cite{van2018widespread} and various frontal areas in Schizophreniform (HC vs SZF) \cite{moon2023magnetic}. This further validates our framework's capability to effectively learn intricate cortical patterns, enabling the identification of novel surface-space anomalies.
\vspace{-5mm}
\section{Conclusion}
\label{sec:conclusion}
We introduced CAM (\underline{C}ortical \underline{A}nomaly detection through \underline{M}asked image modeling), a self-supervised model for unsupervised mental disorder detection using surface features. Our experiments on a psychosis dataset demonstrate that our framework outperformed state-of-the-art methods relying on statistical features. We further identified cortical regions aligning with existing literature, supporting the potential of modeling surface features for identifying novel biomarkers. In the future, we aim to apply CAM in larger, diverse multi-site datasets, considering both hemispheres, and combining it with other modalities. We believe that this approach will provide a novel perspective in elucidating the complex nature of cortical abnormalities in brain disorders and diseases.

\vspace{-2mm}

\section{Compliance with ethical standards}
\label{sec:ethics}
This research study was conducted retrospectively using human subject data from the following sources: 1) The OPN dataset is openly accessible at \url{https://openneuro.org/}. Ethical approval was not required as per the attached open-access data license. 2) The TOP data was obtained from the Thematically Organized Psychosis (TOP) study, which received approval from the Norwegian Regional Committee for Medical Research Ethics and the Norwegian Data Protection Authority.

\vspace{-4mm}
\section{Acknowledgments}
\label{sec:acknowledgments}
We would like to express our gratitude to all participants in this study, whose involvement made this work possible. This work was supported by the German Research Foundation (DFG) Emmy Noether (513851350 (TW)). Additionally, we acknowledge the support from the BMBF-funded de.NBI Cloud within the German Network for Bioinformatics Infrastructure (de.NBI) (031A532B, 031A533A, 031A533B, 031A534A, 031A535A, 031A537A, 031A537B, 031A537C, 031A537D, 031A538A).

\bibliographystyle{IEEEbib}
\vspace{-3.5mm}
\bibliography{strings,refs}

\end{document}